\documentclass[usenatbib]{mn2e}
\usepackage{epsfig}
\usepackage{times}
\usepackage{natbib}
\usepackage{aas_macros}

\title{Evidence for radio-source heating of groups}
\author[J. H. Croston et al.]  {J. H. Croston$^{1,2}$\thanks{Email:
jcroston@discovery.saclay.cea.fr}, M.J. Hardcastle$^{1,3}$ and
M. Birkinshaw$^1$\\$^1$ H. H. Wills Physics Laboratory, University of
Bristol, Tyndall Avenue, Bristol BS8 1TL\\$^2$ Service
d'Astrophysique, CEA Saclay, L'Orme des Merisiers, B\^{a}t. 709, 91191
Gif-sur-Yvette, France\\$^3$ School of Physics, Astronomy and
Mathematics, University of Hertfordshire, College Lane, Hatfield,
Hertfordshire AL10 9AB}
 
\pagerange{\pageref{firstpage}--\pageref{lastpage}}
\pubyear{2004}
\begin{document}

\maketitle

\label{firstpage}

\begin{abstract}

We report evidence that the gas properties of X-ray groups containing
radio galaxies differ from those of radio-quiet groups. For a
well-studied sample of {\it ROSAT}-observed groups, we found that more
than half of the elliptical-dominated groups can be considered
``radio-loud'', and that radio-loud groups are likely to be hotter at
a given X-ray luminosity than radio-quiet groups. We tested three
different models for the origin of the effect and conclude that
radio-source heating is the most likely explanation. We found several
examples of groups where there is strong evidence from {\it Chandra}
or {\it XMM-Newton} images for interactions between the radio source
and the group gas. A variety of radio-source heating processes are
important, including shock-heating by young sources and gentler
heating by larger sources. The heating effects can be longer-lasting
than the radio emission. We show that the sample of X-ray groups used
in our study is not significantly biased in the fraction of radio-loud
groups that it contains. This allows us to conclude that the energy
per particle that low-power radio galaxies can inject over the group
lifetime is comparable to the requirements of structure formation
models.

\end{abstract}


\section{Introduction}

Radio galaxies must be transferring large quantities of energy to the
surrounding group- or cluster-scale gas. The $P$d$V$ work done on the
gas by source expansion is of the order of 10$^{52}$ -- 10$^{53}$ J
for source ages $\sim 10^{8}$ years \citep[e.g.][]{c03b}. The first
strong evidence for radio-source heating was recently found in the
nearest radio galaxy, Centaurus A \citep{kra03}: {\it Chandra}
observations reveal a prominent shell of emission capping the inner
southwestern radio lobe, which has a temperature ten times that of the
surrounding galactic atmosphere, providing strong evidence that Cen A
is shock-heating its atmosphere.  There is also evidence for heating
in the powerful FRII radio galaxy, Cygnus A \citep{smi02}, where a
slight temperature increase in the X-ray gas is seen at the front
edges of the lobe cavities. Another recent result helps to emphasise
that more than one energy transfer mechanism is likely to
operate. Deep {\it Chandra} observations \citep{fab03} of the Perseus
cluster [the first cluster where ``cavities'' were observed
\citep{boh93}] have revealed the presence of ripples in the cluster
gas that are suggestive of sound waves emanating from the central
radio source, 3C~84. Fabian et al. calculate that the time scale
between successive wavefronts is comparable to estimates of the
radio-source lifetime, so that it seems plausible that an intermittent
radio source is producing the observed ripples.

There is evidence that energy injection of some sort is required to
explain the observational properties of X-ray groups and clusters,
which do not agree with the predictions of CDM models of structure
formation \citep[e.g.][]{ae99,pon99}. Several authors have considered
the effects of heating due to feedback from star formation
\citep[e.g.][]{bal99,bm01}. However, in most cases it is found that
heating from star formation and supernovae does not provide sufficient
energy to explain the properties of both groups and clusters
\citep[e.g.][]{kra00}. \citet{wu00} find that supernova heating can
generate only $\sim 1/10$ of the energy required. These results have
led many authors to consider AGN heating.  \citet{chu01} and
\citet{boh02} consider the effects of radio-galaxy heating on cluster
structure and conclude that sufficient heating could be
provided. Evidence from recent observations of clusters, as well as
simulations, suggest that AGN could provide a sufficiently distributed
heating mechanism for this method to work \citep[e.g.][]{bk02,fab03},
which removes a long-standing objection to radio-source heating
models. Since radiative cooling in the absence of feedback
overpredicts the mass in galaxies \citep[e.g.][]{col91}, it seems
likely that some sort of feedback is required to explain observed
cluster properties and the galaxy luminosity function
\citep[e.g.][]{vb01,ben03,kay04}.

In addition, nearly all cooling-flow clusters contain a central radio
galaxy \citep[e.g.][]{eil04}, and there is considerable evidence that
the radio galaxy displaces gas in the cooling-flow regions
\citep[e.g.][]{boh93,bla01,fab03}. It is therefore important to
consider the different means by which radio galaxies could influence
the observed properties of the cooling flow, and in particular whether
they can help explain why large quantities of gas do not appear to
cool past $\sim 1/3$ the outer cluster temperature
\citep[e.g.][]{pet03,sak02}. Although non-heating solutions to this
problem exist, it is most plausible that the bulk of the gas in these
systems is reheated by one of several possible mechanisms, such as
cluster mergers, thermal conduction \citep[e.g.][]{voi02,voi04}, or
AGN heating \citep{bt95,bk01,bk02,rey02}; the last of these is
particularly attractive because of the possibility of self-regulation
via feedback. Investigating how radio galaxies can affect their
hot-gas environments is therefore important to our understanding of
several problems of cluster physics.

In \citet{c03b}, we showed that radio galaxies have an important
impact on groups, and found evidence that the properties of
``radio-loud'' and ``radio-quiet'' groups differ, in the sense that
``radio-loud'' groups are hotter than ``radio-quiet'' groups of
comparable X-ray luminosity. That work used a fairly small and
inhomogeneous sample, and the analysis methods were comparatively
basic. Here we present a detailed analysis of a larger, homogeneous
sample of groups \citep{op04} in order to confirm and investigate
further the conclusions of the earlier work.

\section{Sample selection and analysis}
\label{sec:anal}

\subsection{The elliptical-dominated sample}

This analysis uses the GEMS group sample whose X-ray properties were
presented by \cite{op04} (hereafter OP04). The sample consists of 60
groups, including loose and compact groups that may be spiral- or
elliptical-dominated. For the larger GEMS sample, the $L_{X}/T_{X}$
relation shows more scatter than for the earlier work of \citet{hp00},
so that it may be more difficult to distinguish heating effects. 

We considered only the subset of groups in the OP04 sample that
contain a large elliptical galaxy, as spiral-dominated groups are
unlikely to possess a strong radio galaxy. We found that the scatter
in the $L_{X}/T_{X}$ relation is significantly reduced in our
subsample, suggesting that spiral-dominated groups have different gas
properties from those with a large elliptical galaxy. OP04 did not
report a significant difference in the $L_{X}/T_{X}$ properties of
elliptical- and spiral-dominated groups; however, they do find that
most X-ray bright groups contain a bright central early-type
galaxy. They used the group's spiral fraction to compare group
properties. Our classification on the basis of the dominant galaxy's
morphology may be a more useful measure of the group's history and
current properties. This is supported by OP04's conclusions about the
importance of elliptical brightest group galaxies.

In order to determine whether each group contains a large elliptical
galaxy, we downloaded DSS2 (Digitized Sky
Survey\footnote{http://www.eso.org/dss}) images of each group.
Typically groups were either dominated by one large galaxy, or else
there were several bright galaxies of similar magnitude. In the first
case, we rejected any groups whose dominant galaxy is a spiral or S0.
In the second case, where there was no obviously dominant galaxy, the
group was only rejected if {\it none} of the bright galaxies is an
elliptical. In some cases it was not possible to tell by eye whether a
galaxy has an elliptical or S0 morphology, and so we followed up all
of the elliptical groups using Simbad and NED to confirm the galaxy
morphology. Unfortunately there are many cases where these databases
disagree, and multiple classifications exist in the literature. We
therefore only rejected groups where the ambiguous galaxy was
classified as S0 by both Simbad and NED. In all we excluded 15
spiral-dominated groups: HCG~10, HCG~15, HCG~16, HCG~40, HCG~68,
HCG~92, NGC~1332, NGC~2563, NGC~3227, NGC~3396, NGC~4565, NGC~4725,
NGC~5689, NGC~5907 and NGC~6574. In the course of following up galaxy
morphologies, we also found one group, HCG 22, for which the supposed
group members covered an implausibly large range in redshift. As it is
unclear which galaxies are real members of this group, we excluded it
as well.

In addition to excluding spiral-dominated groups, we also had to
exclude 10 groups for which OP04 could not make X-ray spectral
measurements: HCG~4, HCG~48, HCG~58, NGC~1808, NGC~3640, NGC~3783,
NGC~4151, NGC~4193, NGC~6338 and NGC~7714. As the aim of the study was
to compare the properties of radio-loud and radio-quiet groups, we
also had to exclude two groups with low declinations that are not
covered by the surveys used to identify the radio sources, NGC~1566
and NGC~7144. Finally, NGC~315 was excluded, because \citet{wor03}
find a dominant contribution to the X-ray luminosity from the AGN and
X-ray jet, and HCG~67 was excluded because of ambiguity in whether or
not an identified nearby radio source is actually associated with the
group. The final sample of elliptical-dominated groups contained 30
members. In Table~\ref{sample}, we list the groups in the sample along
with their redshift and the X-ray properties measured by OP04 used in
this work.

\begin{table*}
\begin{footnotesize}
\caption[Groups analysis sample (from OP04)]{The elliptical-dominated groups sample. Properties listed here are taken from OP04. $\beta$-model
parameters marked with a star are estimated using the method described
in the text.}
\label{sample}
\vskip 10pt
\begin{tabular}{lrrrrrrr}
\hline
Group&Redshift&Temperature (keV)&Abundance ($Z_{\sun}$)&log($L_{X}$ ergs$^{-1}$)&$r_{cut}$ (kpc)&$\beta$&$r_{c}$ (kpc)\\
\hline
HCG~42&0.0128&0.75$\pm$0.04&0.29$\pm$0.10&41.99$\pm$0.02&112&0.56&4.69\\
HCG~62&0.0146&1.43$\pm$0.08&2.00$\pm$0.56&43.14$\pm$0.04&282&0.48&2.44\\
HCG~90&0.00880&0.46$\pm$0.06&0.08$\pm$0.03&41.49$\pm$0.05&101&0.41&0.91\\
HCG~97&0.0218&0.82$\pm$0.06&0.23$\pm$0.10&42.37$\pm$0.05&339&0.44&2.73\\
NGC~383&0.0173&1.51$\pm$0.06&0.42$\pm$0.08&43.07$\pm$0.01&633&0.36&2.11\\
NGC~524&0.00793&0.65$\pm$0.07&0.22$\pm$0.15&41.05$\pm$0.05&56&0.45*&0.37*\\
NGC~533&0.0181&1.08$\pm$0.05&0.68$\pm$0.23&42.67$\pm$0.03&372&0.42&2.21\\
NGC~720&0.00587&0.52$\pm$0.03&0.18$\pm$0.02&41.20$\pm$0.02&65&0.47&1.15\\
NGC~741&0.0179&1.21$\pm$0.09&2.00$\pm$0.67&42.44$\pm$0.06&386&0.44&2.30\\
NGC~1052&0.00492&0.41$\pm$0.15&0.00$\pm$0.02&40.08$\pm$0.15&25&0.45*&0.04*\\
NGC~1407&0.00565&1.02$\pm$0.04&0.23$\pm$0.05&41.69$\pm$0.02&105&0.46&0.08\\
NGC~1587&0.0122&0.96$\pm$0.17&0.47$\pm$1.24&41.18$\pm$0.09&77&0.46&4.34\\
NGC~3557&0.00878&0.24$\pm$0.02&0.00$\pm$0.01&42.04$\pm$0.04&95&0.52&1.13\\
NGC~3665&0.00692&0.47$\pm$0.10&0.17$\pm$0.14&41.11$\pm$0.08&71&0.47&1.08\\
NGC~3607&0.00411&0.35$\pm$0.04&0.23$\pm$0.10&41.05$\pm$0.05&62&0.39&1.98\\
NGC~3923&0.00459&0.52$\pm$0.03&0.18$\pm$0.05&40.98$\pm$0.02&34&0.55&0.63\\
NGC~4065&0.0235&1.22$\pm$0.08&0.97$\pm$0.48&42.64$\pm$0.05&425&0.36&3.08\\
NGC~4073&0.0204&1.52$\pm$0.09&0.70$\pm$0.15&43.41$\pm$0.02&470&0.43&9.42\\
NGC~4261&0.00785&1.30$\pm$0.07&1.23$\pm$0.42&41.92$\pm$0.03&112&0.44&40.08\\
NGC~4636&0.00565&0.84$\pm$0.02&0.41$\pm$0.05&41.49$\pm$0.02&68&0.47&0.30\\
NGC~4325&0.0252&0.82$\pm$0.02&0.50$\pm$0.08&43.15$\pm$0.01&307&0.58&27.56\\
NGC~4589&0.00676&0.60$\pm$0.07&0.08$\pm$0.03&41.61$\pm$0.05&122&0.52&9.33\\
NGC~4697&0.00454&0.32$\pm$0.03&0.07$\pm$0.02&41.01$\pm$0.02&53&0.46&1.25\\
NGC~5044&0.00820&1.21$\pm$0.02&0.69$\pm$0.06&43.01$\pm$0.01&180&0.51&5.96\\
NGC~5129&0.0232&0.84$\pm$0.06&0.66$\pm$0.28&42.33$\pm$0.04&151&0.43&3.14\\
NGC~5171&0.0232&1.07$\pm$0.09&1.47$\pm$1.25&42.38$\pm$0.06&298&0.45*&81.26\\
NGC~5322&0.00702&0.23$\pm$0.07&0.00$\pm$0.02&40.71$\pm$0.10&43&0.45*&0.18\\
NGC~5846&0.0063&0.73$\pm$0.02&1.25$\pm$0.69&41.90$\pm$0.02&94&0.51&2.19\\
NGC~5930&0.00969&0.97$\pm$0.27&0.17$\pm$0.12&40.73$\pm$0.07&29&0.45*&0.19*\\
IC~1459&0.00569&0.39$\pm$0.04&0.04$\pm$0.01&41.28$\pm$0.04&121&0.45&0.74\\
\hline
\end{tabular}
\end{footnotesize}
\end{table*}

\subsection{Identifying radio sources}
\label{sec:radiodef}

We then divided the elliptical-dominated groups into radio-loud and
radio-quiet subsamples based on the properties of any radio source
associated with each group. For each group in the sample, we used NED,
NVSS \citep{con98} and FIRST \citep{bec95} to locate radio sources
that could potentially be associated with a group member. We first
checked NED for any known radio galaxies associated with the group's
galaxies. If an associated radio source was found, we adopted the
1.4-GHz radio flux from the NED database. For any groups where a radio
source was not found in this way, we searched for sources within a
radius of 10 arcmin using NVSS and FIRST. We then checked the location
of each candidate for an associated radio source on the DSS2 optical
images to ensure that the radio source was roughly at the centre of a
group galaxy (using SIMBAD to confirm that the galaxy is at the
redshift of the group). If this was not the case, then we rejected the
radio source on the basis that it was probably a background object.
This method should be a more robust way of defining radio-loud and
radio-quiet groups than the method we used in \citet{c03b}, as the
latter may contain spurious ``radio-loud'' designations if any of the
radio sources were not in fact associated with the group.
Table~\ref{rsources} gives the 1.4-GHz radio flux and luminosity
densities, $L_{1.4}$, and the location for all of the associated
sources. We also list the distance between the radio source and group
centre, defined in OP04 as the position of the group-member galaxy
nearest to the centroid of the X-ray emission. In two cases (NGC~5171
and HCG~90) the radio source is a significant distance from the group
centre, which may reduce the likelihood that it could strongly affect
the group gas properties; however, as in both cases the radio-source
is clearly associated with a group galaxy and lies in the extended
X-ray halo detected with {\it ROSAT}, we include these two radio
sources in the sample.

\begin{table*}
\begin{center}
\begin{footnotesize}
\caption[Radio sources associated with groups in the sample]{Radio
  sources associated with groups. Coordinates are for the radio-source
  position. Column 3 ($D$) is the physical distance between the radio
  source and the group centre (see text for discussion). Luminosity densities
  are determined by assuming that the source is at the redshift of the
  group.}
\label{rsources}
\vskip 10pt
\begin{tabular}{lrrrrrr}
\hline
Group&$RA$ (J2000)&$Dec$ (J2000)&$D$/kpc&$F_{1.4}$/Jy&$L_{1.4}$/WHz$^{-1}$&Ref\\
\hline
NGC~383&01 07 24.9&+32 24 45&5.0&0&3.4 $\times 10^{24}$&NED\\
NGC~524&01 24 47.7&+09 32 21.7&0.51&0.0035&5.4 $\times 10^{20}$&NVSS\\
NGC~533&01 25 31.3&+01 45 33&0&0.0291&2.2 $\times 10^{22}$&NED\\
NGC~741&01 56 22.1&+05 37 39.8&6.2&1.0&7.2 $\times 10^{23}$&NVSS\\
NGC~1052&02 41 04.8&-08 15 21.1&0.01&0.913&5.1 $\times 10^{22}$&NVSS\\
NGC~1407&03 40 11.9&-18 34 49.0&0.20&0.0877&6.8 $\times 10^{21}$&NVSS\\
NGC~1587&04 30 39.9&+00 39 42.1&0.22&0.123&4.1 $\times 10^{22}$&NVSS\\
NGC~3557&11 09 57.7&-37 32 20.7&0.91&0.484&1.6 $\times 10^{23}$&NVSS\\
NGC~3607&11 17 01.6&+18 08 44.8&30.0&0.0181&6.4 $\times 10^{20}$&NVSS\\
NGC~3665&11 24 43.4&+38 45 44&0&0.113&1.2 $\times 10^{22}$&NED\\
NGC~3923&11 51 05.0&-28 46 10.5&2.3&0.0312&1.7$\times 10^{21}$&NVSS\\
NGC~4261&12 19 23.2&+05 49 31&0&18.0&2.0 $\times 10^{24}$&NED\\
NGC~4636&12 42 50.3&+02 41 18.7&0.64&0.299&1.3 $\times 10^{22}$&NVSS\\
NGC~5044&13 15 24.0&-16 23 7.6&0.27&0.0347&5.0 $\times 10^{21}$&NVSS\\
NGC~5171&13 29 47.9&+11 42 33.8&186&0.025&3.1 $\times 10^{22}$&NVSS\\
NGC~5930&15 26 06.7&+41 40 21.0&3.7&0.108&1.9 $\times 10^{22}$&NVSS\\
HCG~62&12 53 05.6&-09 12 21&1.7&0.0049&1.6 $\times 10^{21}$&NVSS\\
HCG~90&22 02 02.0&-31 52 10.5&81&0.0368&1.4 $\times 10^{23}$&NVSS\\
IC~1459&22 57 10.7&-36 27 43.0&0.19&1.28&9.1 $\times 10^{22}$&NVSS\\
\hline
\end{tabular}
\end{footnotesize}
\end{center}
\end{table*}

In our earlier analysis \citep{c03b}, we chose a single cut-off to
discriminate between radio-quiet and radio-loud groups. However, at
lower luminosities it is difficult to distinguish between AGN-related
radio emission and emission from other processes, such as
starbursts. The radio source population is dominated by starbursts at
1.4-GHz luminosities below $\sim 10^{23}$ W Hz$^{-1}$
\citep{sad02}. All of the radio source identifications are with
elliptical galaxies, so that a starburst origin is unlikely, and the
majority of the sources have also been previously identified as AGN.
However, we may detect weak AGN emission in elliptical galaxies that
have never had a radio source sufficiently powerful to have affected
the group properties. We therefore tested the importance of the
radio-luminosity cut-off, by carrying out all of the analysis that
follows for three choices. We first used a cut-off ($c0$) of
$L^{cut}_{1.4} = 0$, so that the possession of {\it any} radio source
above the NVSS flux density limit meant that a group was considered to
be radio-loud. We then chose two higher cut-offs, based on the
luminosity density of NGC~3665, a comparatively weak double-lobed
radio galaxy, as in the analysis of \citet{c03b}. These cut-offs ($c1$
and $c2$) are $L^{cut}_{1.4}$ = 1.2 $\times 10^{21}$ W Hz$^{-1}$
(L$_{NGC3665}/10$) and $L^{cut}_{1.4}$ = 6 $\times 10^{21}$ W
Hz$^{-1}$ (L$_{NGC3665}$/2).  Table~\ref{rcuts} gives the total number
of groups in each subsample for the three choices of radio-luminosity
cut-off. We note that the NVSS flux limit of 2.3 mJy may introduce
some bias into the selections, as this corresponds to a limiting
luminosity of 3 $\times 10^{21}$ W Hz$^{-1}$ for the highest redshift
group, NGC~4325, at $z=0.0252$. This limit is close to the cut-off
luminosities, so that for cut-offs $c0$ and $c1$ a few high-redshift
groups could have been incorrectly classed as radio-quiet despite
possessing a radio-source more luminous than the cut-off luminosity.
There are five groups with sufficiently high redshift that a radio
source more luminous than $c1$ could have been missed. However, a
radio source of luminosity $>c2$ would be detectable in all of the
groups, so that the results for this cut-off should provide a check
for whether this bias is important.

\begin{table*}
\begin{center}
\begin{footnotesize}
\caption[Sample sizes for cut-off choices]{Samples sizes for different
  choices of radio-luminosity cut-off, $L^{cut}_{1.4}$ as given in the
  text.}
\label{rcuts}
\vskip 10pt
\begin{tabular}{lrrr}
\hline
Cut-off number&$L_{1.4}$/WHz$^{-1}$&Number in RQ sample&Number in RL sample\\
\hline
$c0$&0&11&19\\
$c1$&1.2 $\times 10^{21}$&13&17\\
$c2$&6 $\times 10^{21}$&16&14\\
\hline
\end{tabular}
\end{footnotesize}
\end{center}
\end{table*}

\begin{figure}
\begin{center}
\epsfig{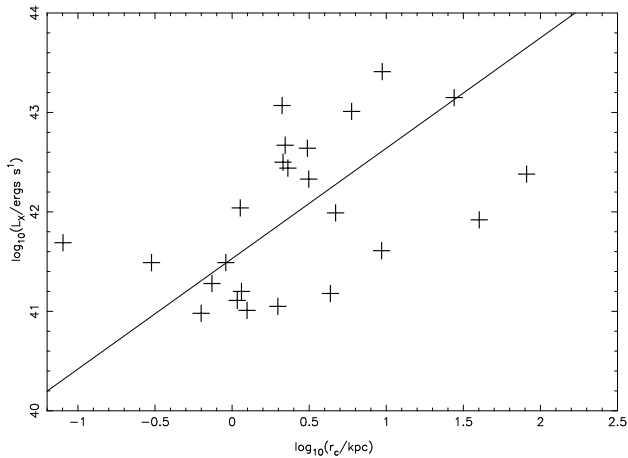}
\caption[$L_{X}/r_{c}$ relation]{Plot of log($L_{X}$) vs.
log($r_{c}$), illustrating the trend between these quantities. The OLS
bisector fit to this line is overplotted and has a slope of 1.11 and
an intercept of 41.53 at $\log10(r_{c}) = 0$. We used this relation to
estimate $r_{c}$ for the groups where OP04 were unable to measure
$\beta$-model parameters.}
\label{lrc}
\end{center}
\end{figure}

\subsection{Radio sources in the OP04 parent population groups}
\label{sec:parentpop}

We found a surprisingly high fraction of radio sources in the sample
of elliptical-dominated groups from the OP04 catalogue (19/30 = 63 per
cent, assuming cut-off $c0$). It is often stated that radio galaxies
are not common; a small fraction of elliptical galaxies [e.g.  $\sim
5$ per cent, \citet{sch78}] host a large radio galaxy. However, the
fraction is certainly higher for the brightest ellipticals
\citep[e.g.][]{bd85}, and \citet{ho99} discusses the recent detections
of small radio cores in many nearby elliptical galaxies, concluding
that these are likely to be low-luminosity AGN. Since the preferred
environment of radio galaxies may be the centres of
elliptical-dominated groups or poor clusters \citep[e.g.][]{bes04},
the high fraction of ``radio-loud'' groups in the sample may not be
unexpected. If radio galaxies are found in such a high fraction of
this type of group, then our results will have important implications
for the properties and evolution of groups. For this reason, it is
crucial to test whether or not the OP04 X-ray observed sample of
groups may be biased in its radio properties. The OP04 sample was
chosen by merging nine catalogues of optical groups and then
cross-correlating the resulting list with the {\it ROSAT} observing
log. The parent catalogue is unbiased with respect to the groups'
radio properties; however, it is possible that the {\it ROSAT} archive
contains a high fraction of groups with active galaxies, since {\it
ROSAT} observed many radio galaxies. This could bias the OP04 sample
towards groups containing radio galaxies, although groups with
previously known radio galaxies make up a fairly small fraction of the
OP04 sample.

To test whether the fraction of radio-loud groups in our sample is
biased, we looked at the parent catalogues used by OP04. As we were
only interested in the properties of elliptical-dominated groups, we
wanted to use an electronically available catalogue that contained
information about the morphology of the dominant galaxy in each group.
The whole-sky group catalogue of \citet{gar93}, taken from the
Lyon-Meudon Extragalactic Database fits these criteria. It contains
485 groups having $z \le 0.02$. This sample is large enough that we
can test whether its radio properties are consistent with those of the
OP04 sample.

Using Vizier\footnote{http://vizier.u-strasbg.fr/}, we extracted from
the Garcia catalogue all groups whose dominant galaxy has type E
(elliptical) or L (S0) (these classifications are taken from
\citealt{dev91}). Although we excluded groups with a dominant galaxy
with a convincing S0 designation in our sample definition, we decided
to include them here, for two reasons. Firstly, as mentioned earlier,
we found several dominant galaxies in our sample with S0 designations
where later work revealed them to be misclassified
ellipticals. Secondly, a surprisingly low fraction of the groups in
the sample had ``E'' designations, so that the test sample would have
been quite small. Including all the S0 groups means that the resulting
radio-loud fraction will be a conservative lower limit, as many of the
S0 identifications will be correct. The final sample of E and S0
groups from the Garcia sample contains 135 groups ($\sim 30$ per cent
of the original sample).

We then cross-correlated the Garcia E and S0 groups with NVSS,
searching for radio sources within 15 arcsec of the centre of the
dominant galaxy. This method is not as accurate as the method we used
for the OP04 radio identifications; however, the more detailed method
is too cumbersome for this larger sample. To ensure a fair comparison,
we carried out the same cross-correlation for our elliptical-dominated
groups sample, using the coordinates given in OP04.

For the E/S0 Garcia subsample, we found radio sources associated with
41 groups. 32 groups had coordinates outside the region covered by
NVSS, so that the final ``radio-loud'' fraction of the Garcia
subsample is 41/103, or 39.8 per cent. For our elliptical-dominated
sample (Table~\ref{sample}), using the same method, we found a
``radio-loud'' fraction of 16/30, or 53.3 percent. Therefore the
``radio-loud'' fraction of the elliptical-dominated groups in OP04 is
consistent with that in this parent catalogue. The fraction we
obtained here for our OP04 subsample is slightly lower than that
obtained using the more accurate identification method described in
Section~\ref{sec:radiodef}, which was 19/30, or 63 percent. This is
unsuprising, as the detailed method would find sources associated with
{\it any} large elliptical in the group, whereas this
cross-correlation method will only find sources associated with the
dominant galaxy.

We conclude that the OP04 group sample is not excessively biased with
respect to the groups' radio properties. The true fraction of
elliptical-dominated groups with radio sources may be $\sim 40 - 50$
percent (since our result for the Garcia sample is a conservative
lower limit), rather than being as high as is suggested by our
original analysis of the OP04 sample. It is interesting that such a
high fraction of elliptical-dominated groups is likely to possess an
AGN-related radio source. If all elliptical-dominated groups are
capable of hosting radio sources, this result could indicate that
radio galaxies have a high duty cycle. Radio sources with obvious
double-lobed structure make up roughly half of the radio sources in
the sample, so that perhaps only half of the 40 - 50 percent of
elliptical-dominated groups with radio sources could be considered to
be in a very active state. Nonetheless, this suggests that a duty
cycle where every elliptical-dominated group contains a radio source
that is active for $\geq$ 1/4 of the time. Weaker sources with no
detected double-lobed structure may be in a less active stage, and the
groups with no radio source above the NVSS limit may be in the least
active phase. If this model is correct, then the effects of radio
sources are likely to be important at some level in all
elliptical-dominated groups. The analysis of the Garcia sample found
that E/S0-dominated groups made up $\sim 30$ percent of the
sample. Although this is a minority of groups, E/S0-dominated groups
are more likely to host a group-scale X-ray atmosphere
\citep{op04}. They are therefore likely to be in a more relaxed state,
so that their X-ray properties will be of more relevance to
structure-formation models.  

\subsection{Issues with X-ray luminosity comparisons}

The analysis we carried out in \citet{c03b} is based on the
$L_{X}/T_{X}$ relations for radio-loud and radio-quiet groups. In that
analysis we did not take into account the choice of radius to which
the X-ray luminosity was measured. OP04 used an X-ray extraction
radius defined by the extent of X-ray emission at a significant level
above the background. They then used the luminosities obtained for
these regions and their fitted $\beta$-model parameters to extrapolate
the luminosity to a fixed overdensity radius, $r_{500}$ (= the radius
corresponding to $500 \times$ the critical density of the
Universe). We were concerned that the choice of radius used for the
luminosity determination might affect the results. There are problems
in this context with both radii used by OP04. A cut-off radius defined
by the extent of X-ray emission biases the X-ray
luminosity-temperature ($L_{X}/T_{X}$) relation in the sense that a
smaller fraction of the atmosphere will be measured for fainter
groups, because the surface brightness drops below the background at a
smaller radius. However, choosing a cut-off at $r_{500}$ may not be
suitable for our analysis either, as the method used by OP04 to define
$r_{500}$ is temperature-dependent, so that the luminosity is measured
to a larger physical radius for hotter groups. If groups have been
heated by the presence of a radio source, the choice of $r_{500}$ will
have the effect of reducing the significance of any heating effect we
measure.

Clearly, it would be preferable to use the luminosity integrated out
to infinity. Unfortunately this is not possible using a $\beta$-model
representation of the groups, as the solid angle integral of the
surface brightness diverges for values of $\beta \leq 1/2$, so that a
cut-off radius must be used for the luminosity extrapolation. As
neither of the two choices of radius used by OP04 is entirely
unbiased, we performed the $L_{X}/T_{X}$ analysis for four choices of
luminosity cut-off radius. These were: the {\it ROSAT} extraction
radius, $r_{cut}$, as given in Table~\ref{sample}; the fixed
over-density radius, r$_{500}$, as calculated by OP04; a physical
radius of 200 kpc, $r_{phys}$; and $r_{4core}$, defined as 4 $\times$
the group core radius, $r_{c}$. We used the luminosity measured to
$r_{cut}$ and the fitted $\beta$-model parameters to calculate the
luminosity to $r_{500}$, $r_{phys}$ and $r_{4core}$. We neglected the
axial ratio parameter, $e$, included in the OP04 fits, which means
that the extrapolated luminosities will be slightly overestimated for
groups that have a large axial ratio. The maximum value of $e$ for a
source in our sample is 2.65 (NGC 4589), but most $e$ values are in
the range of 1.0-1.5.  The radio-loud and radio-quiet groups have
similar distributions of $e$, and so our conclusions should not be
affected. OP04 could not fit a $\beta$ model for five sources, and for
these we used $ \beta$ = 0.45 (the median value measured for the
sample) and $r_{c}$ determined by using the rough correlation between
X-ray luminosity and $r_{c}$ shown in Fig.~\ref{lrc} - $L_{X} = 2.9
\times 10^{41} (r_{c}/$kpc$)^{1.11}$ ergs s$^{-1}$.

Another consideration is whether the measured X-ray luminosities
contain any contribution from sources other than the group gas,
i.e. from AGN and X-ray binaries. {\it Chandra} observations can
resolve the X-ray binary population and allow the integrated
luminosity from X-ray binaries to be determined.  \citet{kra01} find
an integrated X-ray luminosity of $\sim 5 \times 10^{39}$ ergs
s$^{-1}$ for the population of X-ray binaries in Centaurus
A. \citet{zha04} examine the X-ray binary population in NGC~1407, one
of the groups in the sample, which has a comparatively prominent X-ray
binary population, and find that resolved X-ray binaries account for
$<20$ percent of the galaxy-scale extended X-ray emission detected
with {\it Chandra} (or $\sim 1.5 \times 10^{40}$ ergs s$^{-1}$, only
$\sim 3$ per cent of the {\it ROSAT}-measured $L_{X}$ for the
group-scale emission from NGC~1407). We conclude that contamination
from X-ray binaries is not important to our results. Any contamination
from X-ray binary emission that cannot be resolved by {\it Chandra}
should affect the radio-quiet and radio-loud groups in the same
way. We discuss the issue of AGN contamination in
Section~\ref{sec:agncont}.

\subsection{$L_{X}/T_{X}$ relations}
\label{sec:lxtrel}

\begin{figure*}
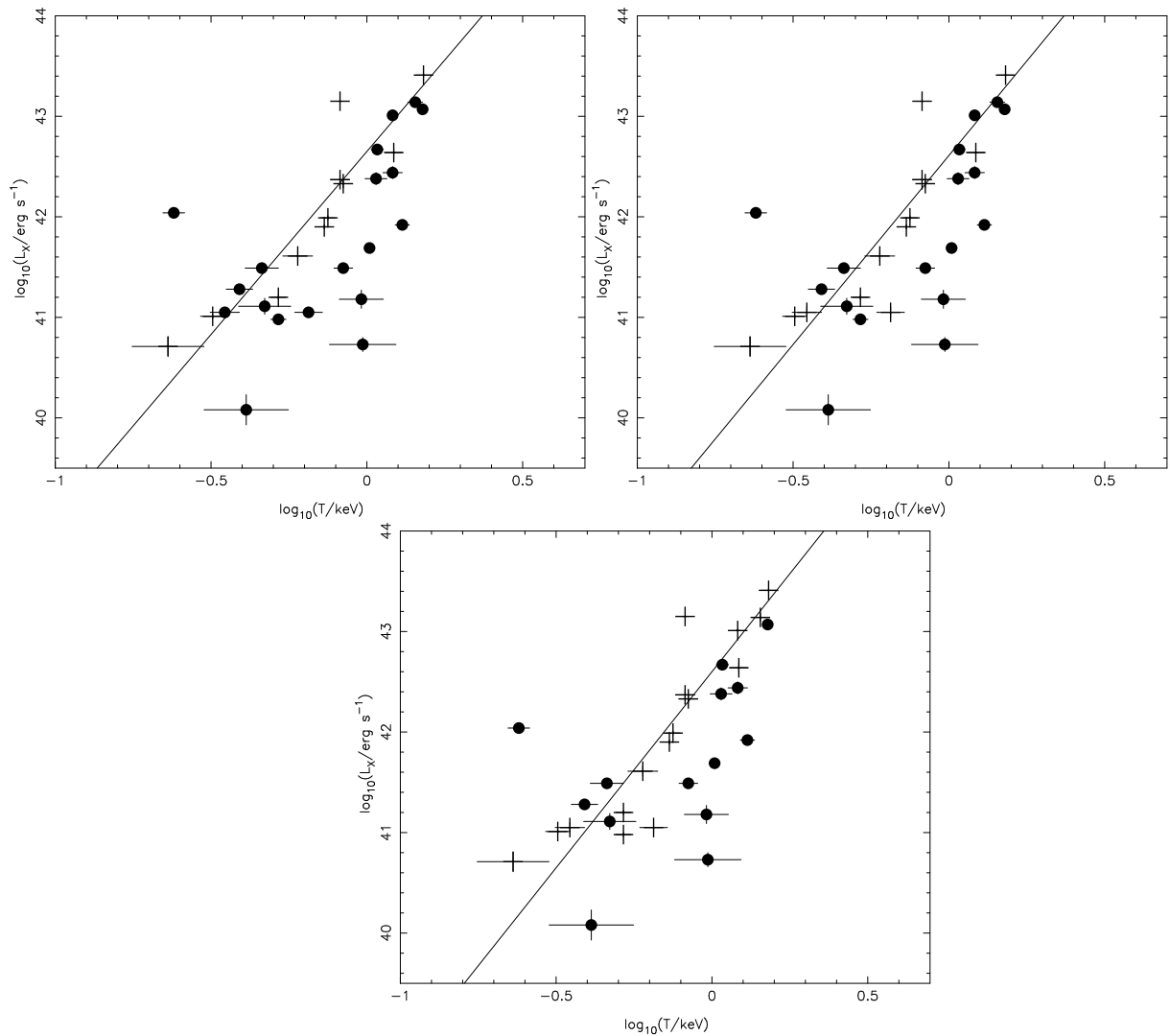

\begin{center}
\centering{\vbox{\hbox{
\epsfig{figure=figures/rcut_c0b.ps,width=8.0cm}
\epsfig{figure=figures/rcut_c1b.ps,width=8.0cm}}
\epsfig{figure=figures/rcut_c2b.ps,width=8.0cm}}}
\caption[$L_{X}/T_{X}$ plots for $c0$]{$L_{X}/T_{X}$ plots for
$r_{cut}$, top left for $c0$, top right $c1$, and bottom $c2$.
Overplotted are the best-fitting radio-quiet relations for each set.
+ symbols are radio-quiet groups and filled circles are radio-loud
groups.}
\label{lt}
\end{center}
\end{figure*}

For each of the 12 combinations of radio and radius cut-offs, we
fitted an $L_{ X}/T_{X}$ relation to the radio-quiet and radio-loud
subsamples. Using the temperature measurements of OP04 and the
luminosities determined as described above, we fitted the OLS
(ordinary least squares) bisector \citep{iso90} to each dataset for
consistency with OP04. Table~\ref{fits} gives the parameters for the
resulting fits. We plot $L_{X}$ vs. $T_{X}$ for each radio cut-off in
Fig.~\ref{lt} (showing only the results for $r_{cut}$) with the
best-fitting radio-quiet relation overplotted. There is a clear
tendency for radio-loud groups to be on the hotter side of the
radio-quiet relation for all combinations of radio cut-off and
luminosity extraction radius. (One major exception to this trend is
NGC~3557, a radio-loud group that is much cooler than the prediction
for its luminosity, but this may be due to including the cooler galaxy
atmospheres of group members, which are prominent in a {\it Chandra}
observation.)

\begin{table*}
\begin{center}
\begin{scriptsize}
\caption[Best-fitting $L_{X}/T_{X}$ relations]{K-S test results and
best-fitting slopes and intercepts for the L$_{X}$/T$_{X}$ relations
as described in the text. $D$ is the K-S statistic for the given pair of radio-quiet and radio-loud samples, and Prob is the null hypothesis probability for obtaining the given value of $D$.}
\label{fits}
\vskip 10pt
\begin{tabular}{lrrrrrrrrrrrr}
\hline
Dataset&\multicolumn{4}{c}{$c0$}&\multicolumn{4}{c}{$c1$}&\multicolumn{4}{c}{$c2$}\\
&Slope&Intercept&$D$&Prob&Slope&Intercept&$D$&Prob&Slope&Intercept&$D$&Prob\\
\hline
$r_{cut}$ (RL)&3.30$\pm0.86$&42.15$\pm0.15$&0.488&0.048&3.33$\pm0.86$&42.16$\pm0.15$&0.416&0.114&2.75$\pm0.71$&42.03$\pm0.18$&0.464&0.054\\
$r_{cut}$ (RQ)&3.64$\pm0.61$&42.65$\pm0.14$&&&3.76$\pm0.72$&42.61$\pm0.14$&&&3.90$\pm0.81$&42.60$\pm0.10$\\
$r_{500}$ (RL)&3.07$\pm0.72$&42.35$\pm0.12$&0.488&0.048&3.19$\pm0.73$&42.35$\pm0.12$&0.434&0.088&2.65$\pm0.59$&42.24$\pm0.14$&0.517&0.022\\
$r_{500}$ (RQ)&3.36$\pm0.51$&42.77$\pm0.12$&&&3.37$\pm0.57$&42.73$\pm0.13$&&&3.58$\pm0.70$&42.73$\pm0.09$\\
$r_{phys}$ (RL)&2.76$\pm0.72$&42.15$\pm0.12$&0.488&0.048&2.84$\pm0.72$&42.15$\pm0.13$&0.452&0.068&2.22$\pm0.51$&42.01$\pm0.12$&0.536&0.016\\
$r_{phys}$ (RQ)&3.05$\pm0.50$&42.57$\pm0.15$&&&3.08$\pm0.56$&42.54$\pm0.15$&&&3.34$\pm0.65$&42.56$\pm0.10$\\
$r_{4core}$ (RL)&3.42$\pm1.10$&41.47$\pm0.18$&0.450&0.084&3.40$\pm1.08$&41.49$\pm0.19$&0.416&0.114&2.68$\pm0.91$&41.29$\pm0.20$&0.518&0.022\\
$r_{4core}$ (RQ)&3.83$\pm0.77$&42.07$\pm0.26$&&&4.02$\pm0.94$&42.02$\pm0.26$&&&4.13$\pm1.04$&42.04$\pm0.18$\\
\hline
\end{tabular}
\end{scriptsize}
\end{center}
\end{table*}

We tested the significance of the trends illustrated in Fig.~\ref{lt}
for each choice of luminosity cut-off. We tranformed the luminosity
values into a predicted temperature using the appropriate best-fitting
radio-quiet $L_{X}/T_{X}$ relation, as given in Table~\ref{fits}.  We
then rotated the coordinate system by $-45$ degrees so that the
$x$-coordinate in temperature corresponds to perpendicular distance
from the best-fitting line. We then performed a 1-D Kolmogorov-Smirnov
test comparing the distributions of $x$ (perpendicular distance from
the line) for the radio-quiet and radio-loud samples of each case. The
results are given in Table~\ref{fits}. In all but two cases, the
probability that the two subsamples have the same parent population is
$<$ 10 percent, and in more than half of the remaining cases, the
probability is less than $<$ 5 percent.

We therefore conclude that there is good evidence that radio-loud and
radio-quiet groups display different gas properties. The choice of
radio-luminosity cut-off does not appear to have an important effect,
whereas the choice of X-ray luminosity radius is crucial. To obtain a
more consistent set of X-ray luminosity and temperature measurements,
higher sensitivity data would be required so that the temperature and
luminosity could be measured to much larger radii, and the need for
extrapolation would be reduced.

\section{Interpretation of the results}
\label{sec:int1}

The gas properties of radio-loud and radio-quiet groups differ in the
sense that radio-loud groups of a given luminosity are likely to be
hotter than the radio-quiet groups. There are several possible
explanations for this result, which we examine in this section.

The first question is whether contamination from AGN emission could
have affected the results for radio-loud groups. We test this in
Section~\ref{sec:agncont}. We then consider three possible physical
origins for the difference in the properties of radio-loud and
radio-quiet groups: radio-source heating (Model I), a luminosity
deficit caused by the radio source (Model II), or an external
mechanism that is triggering the radio source {\it and} heating the
gas (Model III). Models I and II are related, since any temperature
increase would lead to an increase in pressure, expansion and a
subsequent decrease in density on timescales determined by the sound
speed; however, as a first step in understanding the radio-source
impact, it is important to test whether the primary effect seen in our
results comes from an increase in the group temperature or a decrease
in X-ray luminosity. In Section~\ref{sec:rcorrel} we test Model I
(radio-source heating), by examining whether there is any evidence
that the strength of the inferred heating of radio-loud groups
correlates with the properties of the radio sources. In
Sections~\ref{sec:betaprop} and~\ref{sec:lxlbol}, we carry out two
investigations to test Model II (a radio-source-induced luminosity
decrement). Firstly, we examine the distribution of gas, parametrised
by the $\beta$-model, to determine whether this differs for
radio-quiet and radio-loud groups. We then examine the correlation
between $L_{X}$ and optical luminosity to see whether the X-ray
luminosities of radio-loud groups are lower relative to their optical
luminosities than is the case for radio-quiet groups, as would be
expected in the second model. In Section~\ref{sec:opt}, we test Model
III (an external mechanism), by examining the optical properties of
the two subsamples to find out whether the radio-loud groups could be
in a specific evolutionary state, different to that of the radio-quiet
groups, where the triggering of radio sources {\it and} the heating of
gas might be favoured. Finally, in Section~\ref{sec:chgroups} we
present a study of archive {\it Chandra} observations of the
radio-loud groups to look for further evidence of radio-source heating
and interactions between the radio source and group gas.

\subsection{AGN contamination and the reliability of the OP04 results}
\label{sec:agncont}

OP04 state that they have taken into consideration any contribution
from a central AGN to the X-ray emission via their point-source
exclusion method. Central AGN were excluded in 22 cases (Osmond,
private communication); however, this does not include all of the
radio-loud groups in our sample. Contamination from strong non-thermal
emission might result in higher fitted gas temperatures, which could
shift groups with AGN on the $L_{X}/T_{X}$ plane. However, if there
were a large contribution from AGN-related X-ray emission, then the
measured X-ray luminosity from the group gas would be overestimated;
this would act in the opposite sense.  It is therefore essential to
check that AGN contamination is not leading to spuriously high
temperatures or overestimation of the group luminosities, and so we
felt it necessary to confirm the results of OP04 for several
radio-loud groups. In addition, as our results rely strongly on the
accuracy of OP04's {\it ROSAT} analysis, we felt it was appropriate to
independently measure the temperature and luminosity of 3 radio-loud
groups, to test for AGN contamination, and 3 radio-quiet groups, to
confirm the reliability of the luminosity and temperature
measurements.

From the {\it XMM} analysis of \citet{c03b}, we found that the AGN in
3C~66B and 3C~449 (which are more powerful than most of the radio
sources in the sample we use here) would not significantly contaminate
the spectral measurements for the group atmospheres. The measured
temperatures and luminosities in our sample are therefore likely to be
reliable for groups with $L_{X} > 10^{42}$ erg s$^{-1}$. For this
reason, we selected three groups with lower X-ray luminosities:
NGC~4261, NGC~1407, where OP04 did not exclude the central AGN, and
HCG~90.

In order to test the influence of AGN contamination, we extracted a
spectrum for the extraction region used by OP04 (a circle of radius
$r_{cut}$), but excluding the central two arcminutes, so as to ensure
that the majority of the AGN emission was excluded. We used a
surrounding annulus for background and excluded any contaminating
point sources by eye. Our analysis is therefore significantly cruder
than that of OP04, who carried out a more complicated background
estimation and performed good-time interval analysis. In all cases our
measured temperature is the same within the (1$\sigma$)
errors. However, the measured luminosities in two cases are
significantly lower than the OP04 results. We conclude that there is
no risk that the observed difference in the temperature distribution
of radio-quiet and radio-loud groups is due to spuriously high
temperatures as a result of AGN contamination. The lower luminosity
for NGC~4261 is likely to be due to our larger AGN exclusion region
(which must include significant group emission). In the case of
NGC~1407, the slightly lower luminosity is likely to be because OP04
did not exclude the AGN. If the AGN contributes some of the measured
luminosity in a few sources, then the OP04 luminosity measurements for
some of the radio-loud groups may be slightly overestimated, which
would mean that the significance of the effect we observe is
underestimated. If the true errors on temperature for these groups are
slightly larger than those given by OP04, this would not have any effect
on the K-S test results and our conclusions. Finally, since OP04
excluded the AGN in the poorest radio-loud groups, including NGC~1052
and IC~1459, we conclude that AGN contamination is unlikely to be a
problem for our results.

Even if the AGN emission does not significantly affect the measured
temperature, it can have a more important effect on the surface
brightness profile. OP04's neglect of an AGN may affect the fitted
$\beta$-model parameters for a few groups. Several of the most
powerful radio galaxies in the sample (e.g. NGC~383, from which the
AGN was not removed, and NGC~4261) are among the groups fitted with a
second central $\beta$-model. It is possible that these inner
$\beta$-models are, in fact, principally modelling the point-source
emission from the central AGN, although there is also evidence for a
galaxy atmosphere in NGC~383 \citep{h02b}. Since an AGN component
would be modelled out in this way, the $\beta$-model parameters for
the extended emission should be reliable. That a second $\beta$-model
was not required for many of the radio-loud groups, in combination
with the lack of any effect on the measured temperatures, supports the
conclusion that for most of the less powerful radio sources the AGN
contribution to the X-ray emission is not significant.

The radio-quiet groups chosen to check the reliability of the OP04
results were selected to cover a wide range in X-ray luminosity and
temperature. They are NGC~97, NGC~720, and NGC~4325. For each group,
we extracted a spectrum using the extraction region of OP04 (a circle
of radius $r_{cut}$) using the same background and point-source
identification methods as above, and fitted a {\it mekal} model with
free abundance to determine the temperature and X-ray flux of the
group atmosphere. In all cases the results are in good agreement with
those of OP04, so we conclude that the OP04 luminosity and temperature
determinations are reliable both for radio-quiet and radio-loud
groups.

\subsection{Testing Model I: correlations with radio luminosity}
\label{sec:rcorrel}

The results presented in Section~\ref{sec:anal} strongly suggest that
radio galaxies are having an important effect on the properties of the
surrounding group gas. We therefore decided to investigate whether
there is any relationship between the observed ``temperature
excesses'' and the radio properties of the associated sources, as
would be expected in a radio-source heating model. In the following
analysis, we use only the results for radio-luminosity cut-off $c0$ so
as to include the widest range of radio powers.

We used the 1.4-GHz radio luminosity density (Table~\ref{rsources}) as
a measure of the amount of energy a given source might be able to
provide. Radio luminosity is not an ideal indicator of radio source
energy input, as the amount of energy transferred from the expanding
radio plasma to the surrounding gas depends on several factors, such
as source size and age, which are in many cases unknown. The size of
the source depends on its angle to the line of sight, which is usually
poorly constrained. The source's age, which is also required to
estimate the total energy input, is even more difficult to determine.
Indeed, a few of the radio sources in the sample are unresolved with
NVSS and FIRST, and do not have identifiable double-lobed
structure. However, low-frequency radio luminosity should trace the
jet kinetic luminosity reasonably well, so that in the absence of
useful information on the sizes and ages of all of the sources, the
1.4-GHz luminosity is the best measure of energy input available.

We first compared the radio luminosity with $\Delta T$, the difference
between the measured temperature and that predicted by the appropriate
RQ $L_{X}/T_{X}$ relation. Table~\ref{spearman} gives the results of
Spearman rank correlation tests for each choice of X-ray luminosity
radius. For the Spearman test, groups with no temperature increase
were assigned $\Delta T=0$: this applies to 5/19 groups for $r_{cut}$
and $r_{500}$, 6/19 for $r_{phys}$, and 3/19 for $r_{core}$. In all
cases there is little evidence for a correlation. This is perhaps
unsurprising, because the observed temperature increase produced by a
radio galaxy of given luminosity should depend not only on the unknown
properties of the radio source, as mentioned above, but also on the
heat capacity of the group gas being heated. A similar analysis using
the fractional temperature change, $\Delta T/ T$, did not give an
improved correlation.

We therefore estimated the heat capacity of each group's environment
using the spectral and spatial properties of the X-ray emission given
by OP04. The heat capacity is $C = (3/2) N k$, where $N$ is the total
number of particles, obtained by integrating over the density profile
using the best-fitting $\beta$-model parameters, with a central proton
density obtained from the X-ray luminosity and best-fitting {\it
mekal} model parameters.

\begin{figure*}
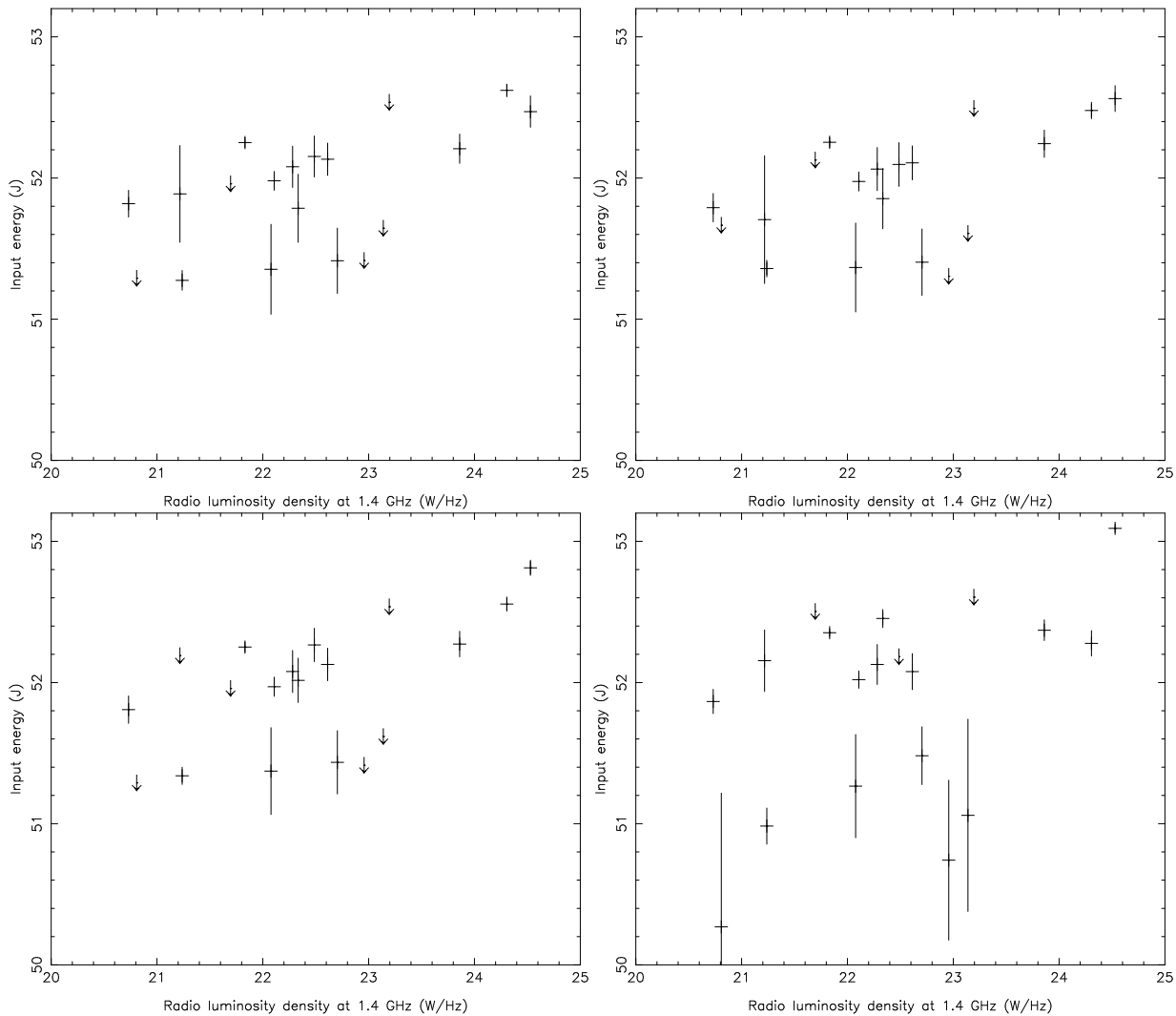

\centering{\vbox{\hbox{
\epsfig{figure=figures/rcut_hc2.ps,width=7.0cm,angle=270}
\epsfig{figure=figures/r500_hc2.ps,width=7.0cm,angle=270}}
\hbox{
\epsfig{figure=figures/rphys_hc2.ps,width=7.0cm,angle=270}
\epsfig{figure=figures/rcore_hc2.ps,width=7.0cm,angle=270}}}}
\caption[$L_{1.4}/ E_{req}$ correlations]{Plot of $L_{1.4}$, the
  1.4-GHz radio luminosity, vs. $E_{req}$, the necessary heat in put
  to produce the temperature increase, for $r_{cut}$ (top left),
  $r_{500}$ (top right), $r_{ phys}$ (bottom left) and $r_{4core}$
  (bottom right). + symbols are groups with temperature excesses;
  arrows indicate upper limits for groups with no observed excess,
  calculated as described in the text.}
\label{eplot}
\end{figure*}

To study the relationship between the observed ``heating'' and radio
luminosity, we examined the correlation between $E_{req}$, the energy
required to heat the gas in a given group from the predicted
temperature to the measured temperature ($C\Delta T$), and $L_{1.4}$.
The heat capacities were calculated separately for each of the four
choices of limiting radius. Fig.~\ref{eplot} shows the relationship
between $L_{1.4}$ and $E_{req}$ for each choice of $r$. For groups
with no temperature excess, we calculated an upper limit to the energy
input by determining the amount of energy that would be required to
shift the group significantly to the `hotter' side of the appropriate
$L_{X}/T_{X}$ relation. As the sample includes upper limits, we used
survival analysis techniques to determine the generalised Kendall's
$\tau$ correlation coefficient using {\sc asurv} \citep{lav92}.
Table~\ref{spearman} contains the results of the correlation analysis
for each case. There is a less than 5 percent probability of obtaining
the measured value of $\tau$ by chance for 2 out of 4 cases. The high
value of 14 per cent for $r_{4core}$ is probably because in many cases
$r_{4core}$ is physically small compared to the other choices for $r$,
so that the heat capacity does not include much of the gas. For all
four choices of radius, there is a stronger correlation here than was
found for $\Delta T$ alone.

\begin{table*}
\begin{center}
\begin{footnotesize}
\caption[Spearman rank tests for $L_{1.4}$ and $\Delta T$, $E_{req}$
  and $C$]{Results of correlation analysis for $L_{1.4}$ and $\Delta
  T$,$E_{req}$ and $C$. The sample contains 19 groups, so for all of
  the tests there are 17 degrees of freedom.}
\label{spearman}
\vskip 10pt
\begin{tabular}{lrrrrrr}
\hline
Dataset&\multicolumn{2}{c}{$\Delta T$}&\multicolumn{2}{c}{$E_{req}$}&\multicolumn{2}{c}{$C$}\\
&$r_{s}$&Probability&Kendall's $\tau$&Probability&$r_{s}$&Probability\\
\hline
$r_{cut}$&0.184&0.450&1.992&0.046&0.312&0.193\\
$r_{500}$&0.214&0.379&1.921&0.055&0.325&0.175\\
$r_{phys}$&0.369&0.121&2.555&0.011&0.312&0.193\\
$r_{4core}$&0.168&0.491&1.472&0.141&0.314&0.190\\
\hline
\end{tabular}
\end{footnotesize}
\end{center}
\end{table*}

Since the calculated heat capacity is related to the measured X-ray
luminosity, we were concerned that the correlation between $L_{1.4}$
and $E_{req}$ could be caused by an $L_{X}/L_{1.4}$ correlation due to
the flux limits in the X-ray and radio samples. We therefore also
carried out Spearman rank tests to look for a correlation between heat
capacity and $L_{1.4}$. Those results are also included in
Table~\ref{spearman}, and show that the correlation between $L_{1.4}$
and heat capacity is much weaker than that with $E_{req}$ in all
cases.

The presence of a correlation (albeit with a large scatter) between
radio luminosity and the energy input needed to cause the observed
temperature increase provides support for a model where the
temperature increase is due to radio-source heating. The large scatter
is not surprising, given the many unknown factors, such as source size
and age, that would affect the amount of observed heating.

\subsection{Testing Model II: $\beta$-model properties}
\label{sec:betaprop}

\begin{figure*}
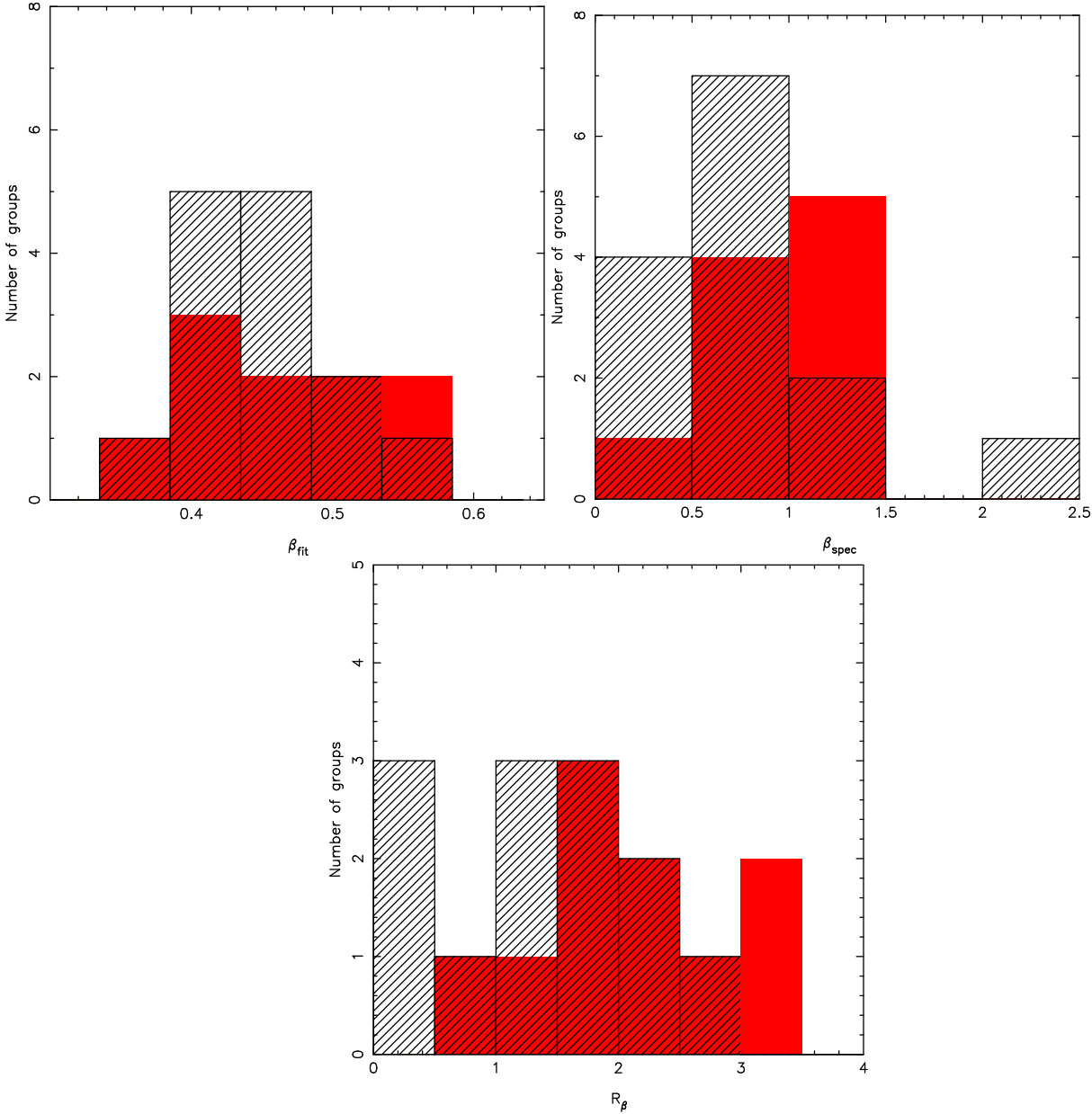

\centering{\vbox{\hbox{
\epsfig{figure=figures/betafit.ps,width=8.0cm}
\epsfig{figure=figures/betaspec.ps,width=8.0cm}}
\epsfig{figure=figures/betaratio.ps,width=8.0cm}}}
\caption[Histograms of $\beta_{fit}$, $\beta_{spec}$ and
  $R_{\beta}$]{Histograms showing the distribution of $\beta_{fit}$
  (top left), $\beta_{spec}$ (top right), and $R_{\beta}$ (bottom) for
  the sample. Filled rectangles are the radio-quiet sample, with the
  radio-loud sample overplotted as hatched rectangles.}
\label{betahists}
\end{figure*}

We studied the $\beta$-model properties of the radio-quiet and
radio-loud groups to determine whether the spatial distribution of gas
is affected by the presence of a radio source. We compared three
parameters, $\beta_{fit}$, the fitted value of $\beta$ from OP04,
$\beta_{spec}$, the spectroscopic value of $\beta$ from OP04, and
$R_{\beta} = \beta_{spec}/\beta_{fit}$. For each of these properties,
we compared the values for the radio-quiet and radio-loud subsamples
for radio cut-off $c0$. We performed a 1-D K-S test to determine
whether the ``radio-quiet'' and ``radio-loud'' subsamples differed in
each case. Fig.~\ref{betahists} shows histograms of the distributions
of $\beta_{fit}$, $\beta_{spec}$, and $R_{\beta}$ for the radio-quiet
and radio-loud samples.

There is no evidence that the parent population differs significantly
for any of the three parameters. Since the $\beta_{spec}$ have large
errors, using the K-S test to compare the two samples may not be very
reliable. We also used a median test to compare the two samples, and
find a low probability that the distributions of $\beta_{fit}$ have a
different median. However, there is a probability of $\sim 93$ percent
that the distributions of both $\beta_{spec}$ and $R_{\beta}$ have
different medians for radio-quiet and radio-loud groups, in the sense
that $\beta_{spec}$ is higher for RQ groups. This is not a strong
result, because of the large errors on $\beta_{spec}$ and therefore
$R_{\beta}$. There is no evidence that RL groups have flatter profiles
than RQ groups, as might be expected if the luminosity had been
significantly decreased as a result of radio-galaxy input. In
Section~\ref{sec:interp} we present further discussion of how the
group density distribution might be affected by radio-galaxy energy
input

\subsection{Testing Model II: $L_{X}/L_{B}$ relation}
\label{sec:lxlbol}

\begin{figure}
\begin{center}
\epsfig{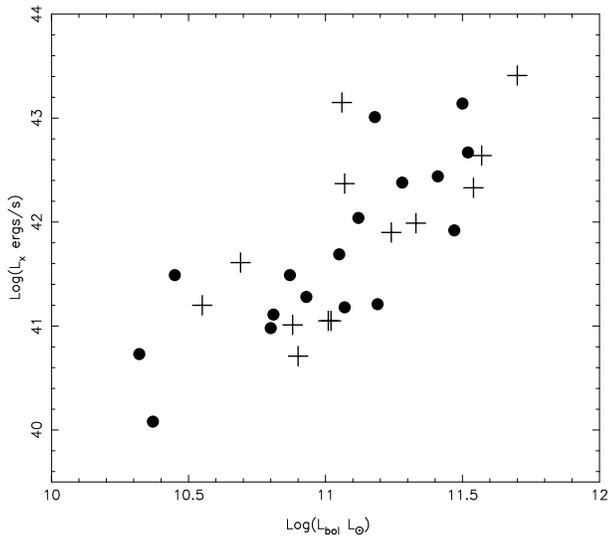}
\caption[$L_{X}/L_{B}$ relation for radio-quiet and radio-loud
  groups]{The $L_{X}/L_{B}$ relation for radio-quiet and radio-loud
  groups. Symbols are as for previous figures.}
\label{lxlbol}
\end{center}
\end{figure}

The X-ray and optical luminosities of groups are correlated, because
gas mass and galaxy mass should scale similarly. OP04 show that such a
correlation exists for their sample. If the effects we observe in
Fig.~\ref{lt} are caused by a decrease in X-ray luminosity in the
radio-loud groups, then the $L_{X}/L_{B}$ relation should be affected:
radio-loud groups should have a lower X-ray luminosity relative to
their optical luminosity. In Fig.~\ref{lxlbol}, we show the
$L_{X}/L_{B}$ relation for radio-quiet and radio-loud groups (using
$c1$ and $r_{cut}$). Unlike what is seen for the $L_{X}/T_{X}$
relation, there is no apparent difference in the two subsamples. We
note that that the radio luminosity will be related to $L_{B}$, which
may introduce a slight bias, but this should not affect the
X-ray-to-optical luminosity ratio of the radio-loud groups. Therefore,
this is a strong argument against X-ray luminosity decrements in
radio-loud groups, as the radio source should not affect the optical
luminosity of the group.

\subsection{Testing Model III: optical properties of the RL and RQ subsamples}
\label{sec:opt}

\begin{figure*}
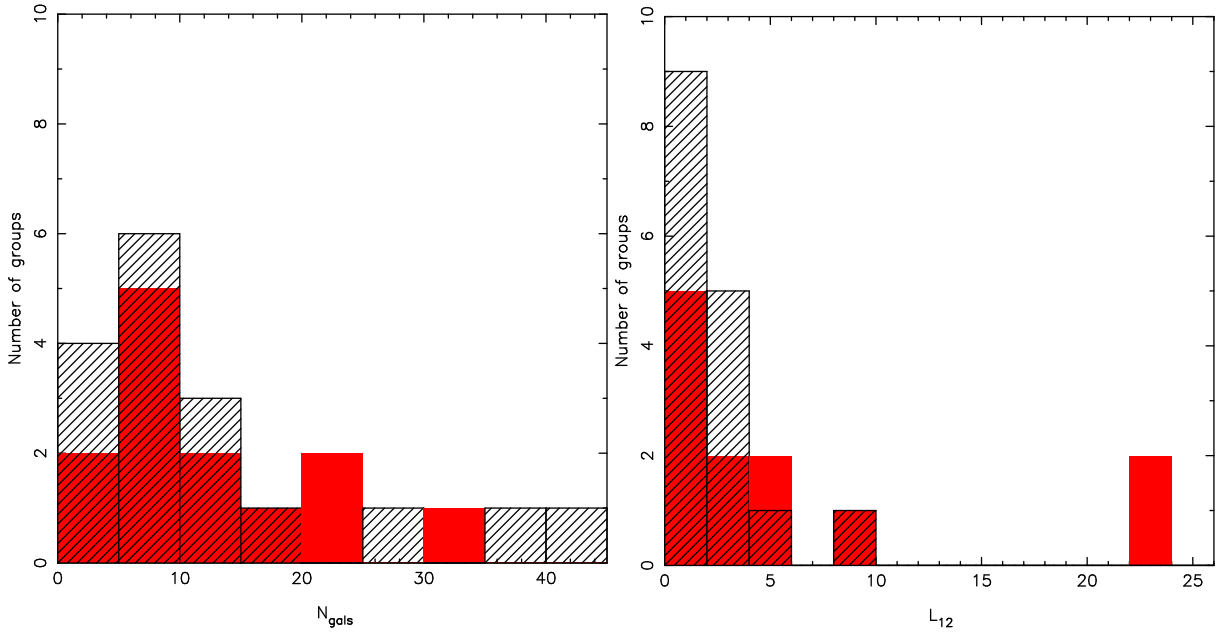

\centering{\hbox{
\epsfig{figure=figures/ngalshist.ps,width=8.0cm}
\epsfig{figure=figures/groupshist.ps,width=8.0cm}}}
\caption[Histograms of $N_{gals}$ and $L_{12}$]{Histograms showing the
  distribution of $N_{gals}$ (left) and $L_{12}$ (right) for the
  radio-quiet and radio-loud samples. The RQ sample is indicated by
  filled rectangles, and the RL distribution is overplotted with
  hatched rectangles in both plots.}
\label{hists}
\end{figure*}

To test the possibility that radio-quiet and radio-loud groups are in
different stages of evolution, so that an external mechanism might be
causing the heating effect, we compared two measures of their optical
properties.  If radio-quiet groups are at a different stage of
evolution from radio-loud groups, then $N_{gals}$, the number of
galaxies in the group, might be expected to differ, in the sense that
older groups might be expected to have fewer members.  Older groups
might also have a larger ratio between the brightest and
second-brightest group galaxies, as the largest mergers should have
already taken place.  We therefore compared both $N_{gals}$ and
$L_{12}$, the luminosity ratio of the brightest to second-brightest
group galaxy, as determined by OP04, for the two subsamples (using $c1$
and $r_{cut}$).  In Fig.~\ref{hists}, we show histograms of the
distribution of these parameters for the two subsamples. In neither
case is there a significant difference in the distribution for the two
subsamples. Both have a peak in the value of $N_{gals}$ between 5 and
10, and the preferred value of $L_{12}$ for both subsamples is $<2$,
so that most groups have at least one reasonably large secondary
galaxy, whether radio-loud or not. There are no radio-loud groups in
the sample with $L_{12} > 10$, whereas there are two radio-quiet
groups with $L_{12} > 20$.  However, a K-S test indicates no
significant difference in the distributions of either. A median test
also finds no significant difference in the medians of the two
subsamples for either parameter. We conclude that the galaxy
distributions in radio-quiet and radio-loud elliptical-dominated
groups are similar, so that there is no evidence that the two
subsamples are at different evolutionary stages. However, a thorough
investigation using more sophisticated measures of group history is
required to test this model fully.

\subsection{{\it Chandra} and {\it XMM-Newton} observations of heating and interactions
  in the RL groups}
\label{sec:chgroups}

Many of the radio-loud groups in this sample have been observed with
{\it Chandra} or {\it XMM-Newton}. The high resolution of {\it
Chandra} is excellent for resolving an AGN component, and for
detecting inner structure in groups; however, as a result of the high
resolution, its sensitivity to extended emission is reduced, so that
in some cases it is unable to detect low surface-brightness emission
from the weaker groups. In those cases {\it ROSAT} temperature and
luminosity measurements are likely to be superior. We discuss here a
few groups in the sample for which {\it Chandra} and {\it XMM-Newton}
observations show evidence for heating or interactions between a
radio-source and its environment.

\subsubsection{NGC~4261}

\begin{figure*}
\begin{center}
\epsfig{figure=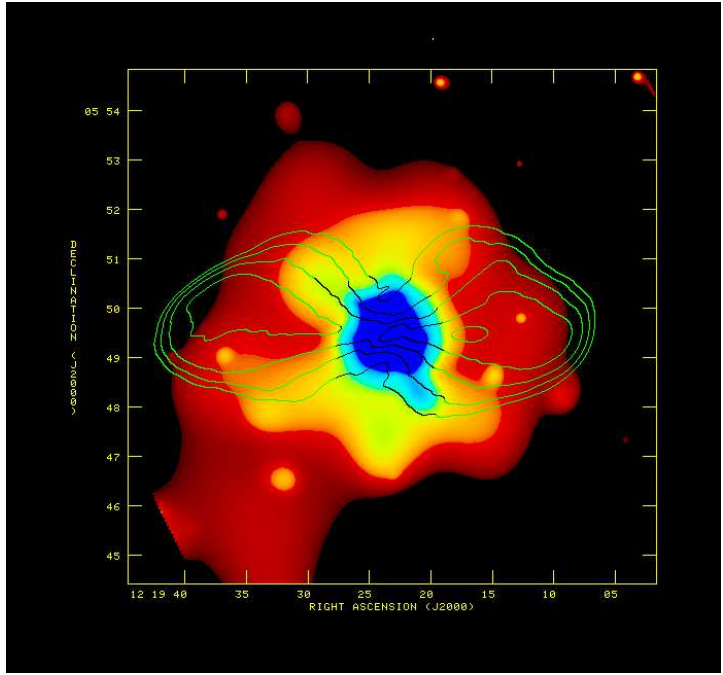,width=9.5cm}
\caption[{\it XMM-Newton} image of NGC~4261 with radio contours
  overlaid]{An adaptively smoothed, background point-source
  subtracted, vignetting-corrected 0.5 -- 5.0 keV image of NGC~4261
  made from the combined MOS1, MOS2 and pn data from the archive {\it
  XMM} observation described in the text. Clear evidence for
  interactions between the radio galaxy (3C~270) and gas environment
  are seen in the form of holes in the X-ray surface brightness at the
  positions of both radio lobes.}
\label{xmm4261}
\end{center}
\end{figure*}

An {\it XMM-Newton} observation of this group was made on 16 December
2001 (ObsID 0056340101). An analysis of the extended emission has not
yet been been published, although \citet{sam02b} presented an analysis
of the nuclear emission. We extracted the archive {\it XMM-Newton}
data and carried out standard processing and filtering as described in
\citet{c03b}. Fig.~\ref{xmm4261} shows the adaptively smoothed,
background point-source subtracted, vignetting-corrected 0.5 -- 5.0
keV image made from combined MOS1, MOS2 and pn data, with radio
contours from a 1.4-GHz map made from VLA archive data overlaid. This
figure illustrates a striking relationship between radio and X-ray
morphology similar to that seen in 3C~66B \citep{c03b}. It is
interesting that such evidence for interactions between the radio
source and hot gas on large scales is found in every FRI radio galaxy
for which deep {\it XMM} images of the large-scale structure exist
[see also \citet{eva04}].

\subsubsection{NGC~4636}

The {\it Chandra} observation of the atmosphere surrounding NGC~4636
revealed striking substructure in the form of symmetrical bright arms
\citep{jon02}. Jones et al. find that the leading edges of the arms
are $\sim 30$ percent hotter than the surrounding gas, and postulate a
model in which the arms are produced by shocks driven by symmetric
off-centre AGN outbursts. The central radio source appears to be
extended to the northeast and southwest \citep[e.g.][]{bd85}; however,
it is small and too weak to have produced the shock-heating. Jones et
al. argue that this indicates that a more direct nuclear outburst may
have produced the shocks, but it is difficult to think of such a
mechanism. The radio source could be more extended at low frequencies;
however, it remains unlikely that a currently active radio source is
producing the shocks.  It is possible that a previous radio outburst
is responsible, although it is unclear for how long the sharp density
and temperature structure would persist. The shock-heated arms of gas
in this group may be the main contributor to the overall temperature
of 0.84 keV measured by {\it ROSAT}. As NGC~4636 has one of the
largest temperature excesses, this example strongly suggests that we
are indeed identifying groups with interesting AGN/group interactions.

\subsubsection{NGC~1052}
\label{sec:1052}

\begin{figure}
\begin{center}
\epsfig{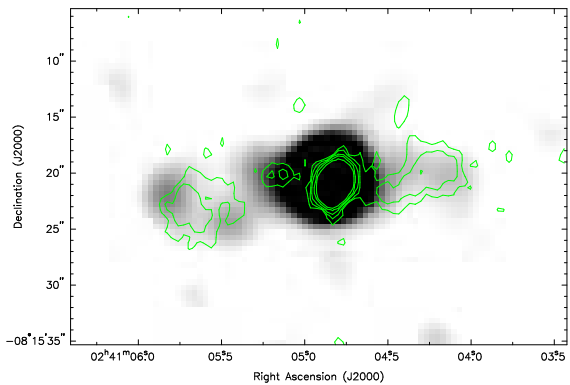}
\caption[Radio contours overlaid on {\it Chandra} image of
  NGC~1052]{Gaussian smoothed image of the {\it Chandra} data for
  NGC~1052 in the energy range 0.5 -- 5.0 keV. Radio contours are
  overlaid from a map made from archive VLA data \citep{wro84}.}
\label{group1052}
\end{center}
\end{figure}

The {\it Chandra} observation is too short to detect much low
surface-brightness emission from this poor group. However, as shown in
Fig.~\ref{group1052}, there is evidence for radio-related X-ray
emission, as discussed by \citet{kad04}. They attribute most of the
X-ray emission to the jet; however, the distribution of X-ray counts
around the eastern radio lobe seems to be reminiscent of the bright
shell of hot gas around the southwestern inner lobe of Centaurus
A. The two systems are remarkably similar: they both consist of small,
double-lobed radio sources, which are likely to be young or
restarting. They have galaxy atmospheres of similar X-ray luminosities
and temperatures. These {\it Chandra} observations therefore suggest
that the young radio source in NGC~1052 may be shock-heating its
surroundings. The measured X-ray temperature from {\it ROSAT} may
contain a large contribution from these radio-related regions,
although it is also possible that the entire environment has been
heated, as we have argued to be the case for 3C~66B \citep{c03b}.

\subsubsection{HCG~62}

The X-ray emission from HCG~62 provides one of the clearest examples
of ``holes'' in a group atmosphere \citep{vri00}. However, the current
AGN is a weak radio emitter, and does not show any extension. As with
NGC~4636, it is plausible that a previous radio outburst (which may
still be detectable in low-frequency radio observations) has produced
the observed X-ray structure.

\subsubsection{NGC~5044}

This group shows prominent substructure in the {\it Chandra} image of
\citet{buo03}, who associate holes in the gas with the radio
source. Although the NVSS image of this source does not show evidence
for any extended radio emission, they suggest that observations at
lower frequency might reveal the presence of radio emission filling
the cavities. We have been unable to resolve the AGN or detect
larger-scale radio emission in our analysis of VLA archive data.

\subsubsection{Summary of the {\it Chandra} and {\it XMM-Newton} observations}
\label{sec:chsum}

The {\it Chandra} and {\it XMM-Newton} observations discussed above
show that radio-source/group interactions are complex. In two cases,
NGC~1052 and NGC~4261, there is clear evidence that the X-ray
structure has been affected by the current radio galaxies. In two
groups where there are no detected large-scale radio lobes, HCG~62 and
NGC~4636, the {\it Chandra} observations reveal striking X-ray
morphologies suggestive of outbursts from the AGN. Smaller-scale
substructure is also present in NGC~5044. Finally, localised heating
appears to be present in NGC~1052 and NGC~4636, suggesting that the
heating effects we observe from the {\it ROSAT} sample could be caused
by several different processes.

\section{Evidence for radio-source heating?}
\label{sec:interp}

The results presented in Section~\ref{sec:anal} support the argument
that radio galaxies have an important effect on the X-ray properties,
and therefore the physical conditions, of group gas, as suggested by
previous work \citep[e.g.][]{c03b}. This is shown by the difference in
$L_{X}/T_{X}$ properties of the two subsamples, and by the high
incidence of radio-related substructure in radio-loud groups.

In the next two sections we discuss the three models of
Section~\ref{sec:int1} in detail. In Section~\ref{sec:tinc}, we
compare models I and II, and argue in favour of a radio-source heating
model (Model I) and against a luminosity deficit (Model II). In
Section~\ref{sec:expl3}, we consider one possible external mechanism
that could lead to a Model III explanation, that of mergers and
interactions, and argue against such a model. In
Section~\ref{sec:models}, we attempt to explain the results for all of
the radio-loud groups in the context of a model of radio-source
heating and discuss what can be inferred about the heating processes.

\subsection{Temperature increase vs. luminosity decrement}
\label{sec:tinc}

Radio galaxies must displace large amounts of gas and this could have
a significant effect on their luminosity. For 3C~66B, which is larger
than most of the sources in the samples studied here, we calculated
that the gas with which the radio source can have directly interacted
provides only 7 percent of the group's luminosity \citep{c03b}. It is
therefore unlikely that removal of gas by the radio galaxy could
produce the luminosity deficits needed by this model, in some cases an
order of magnitude in luminosity. However, the group luminosity will
also be decreased if a significant fraction of the jet kinetic energy
is transferred into potential energy.

In the context of preheating models of energy injection into group
gas, it has been argued \citep[e.g.][]{me94,hp00} that the main effect
of the energy injection will be an increase in the group's potential
energy, so that the central density decreases (and hence luminosity
will decrease). While (by the virial theorem) this will eventually be
the case, heating effects are still likely to be detectable on shorter
timescales. More recently, \citet{kay04} has carried out cosmological
simulations of cluster formation including cooling and feedback (which
could be due to AGN or a different energy source such as supernova
winds) and find gas properties in agreement with observations, with a
$\sim 10$ percent increase in temperature at the virial radius. Their
simulations consider only massive clusters, but suggest that at least
some fraction of AGN-injected energy is likely to end up in the
thermal energy of the group.

In some of the groups in our sample an order of magnitude decrease in
luminosity is required: the density would have to be dramatically
reduced to produce such an effect. A strong argument against such
large luminosity deficits in radio-loud groups comes from the
$L_{X}/L_{B}$ relation discussed in Section~\ref{sec:lxlbol}. We find
that the radio-loud groups follow the same trend as the radio-quiet
groups and show no evidence for having lower X-ray luminosities
relative to their optical luminosities, as would be expected if an
X-ray luminosity decrease had been caused by the radio galaxy.

Another strong argument in favour of heating as the dominant effect,
as opposed to a change in luminosity, is the result of
Section~\ref{sec:rcorrel}, where we found evidence for a correlation
between radio luminosity and the energy required to heat the gas from
the predicted to the observed temperature. This result would be harder
to explain in a model where the radio-source's impact was principally
on the group's luminosity.

\begin{figure}
\begin{center}
\epsfig{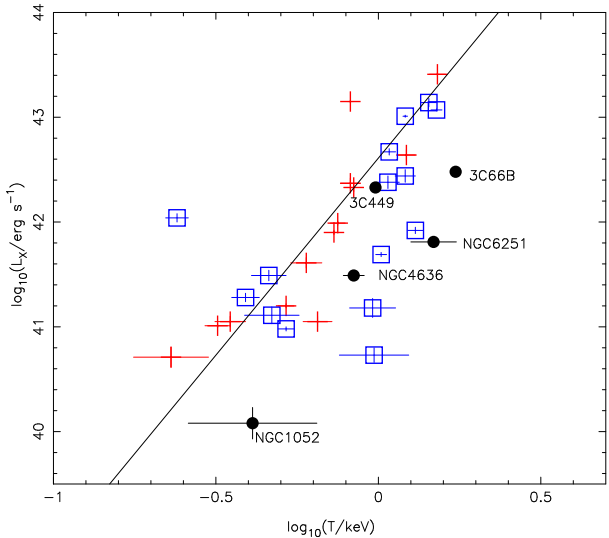}
\caption[$L_{X}/T_{X}$ relation showing the {\it XMM}-observed groups
  and those in the sample with additional evidence for
  heating]{$L_{X}/T_{X}$ relation ($c1$, $r_{cut}$) showing the
  positions of the three {\it XMM}-observed radio-galaxy environments
  presented in \citet{c03b} and \citet{eva04}, as well as the two
  OP04 groups with additional strong evidence for heating, NGC~1052 and
  NGC~4636, marked. + symbols indicate radio-quiet groups, and hollow
  squares radio-loud groups.}
\label{ltannot}
\end{center}
\end{figure}

As shown in Section~\ref{sec:chgroups}, several groups with a
temperature excess possess additional evidence for radio-source
heating. In Fig.~\ref{ltannot}, we show the $L_{X}/T_{X}$ relation for
$c1$ and $r_{cut}$, with NGC~4636 and NGC~1052, as well as 3C~66B,
3C~449 and NGC~6251 \citep{c03b,eva04} marked to illustrate how they
compare to the sample studied here. We conclude, based on the
additional evidence for heating in several sources, and the arguments
above, that heating is a more plausible explanation than a
radio-source induced luminosity deficit.

On the longest timescales, the energy injected by radio galaxies into
group or cluster gas must predominantly end up as potential energy,
and any long-term temperature increase will be small. However,
information about the energy injection cannot travel faster than the
sound speed, so that temperature effects may be detectable for a few
sound-crossing times. That an $L_{X}/T_{X}$ relation for
elliptical-dominated groups exists at all is evidence that, on
average, the temperature increase must disappear on timescales less
than the radio-source recurrence time; for a 50 per cent radio-galaxy
duty cycle, this is comparable to a few sound crossing times in a
typical group. It is plausible that occasionally a second AGN outburst
would occur before the group has recovered from the previous outburst,
so that the temperature increase is more persistent; 3C~66B could be
one example of such a system. Our results therefore suggest that we are
detecting the short-term effects of radio-source heating in many
elliptical-dominated groups. The fact that we have found no systematic
differences in the properties of radio-loud and radio-quiet groups,
other than their $L_{X}/T_{X}$ relations, is consistent with a model
in which all elliptical-dominated groups have had similar numbers of
radio-galaxy outbursts averaged over the group lifetime, affecting the
groups by causing a temporary increase in temperature, with less
easily detectable long-term effects on the group luminosity and
surface brightness distribution.

\subsection{The evolution of groups: a common cause for heating and
  radio activity?}
\label{sec:expl3}

An alternative explanation is that some common cause triggers a radio
source and heats the gas. It is possible that the elliptical-dominated
groups containing radio sources exist at a particular stage in the
evolutionary process for groups, where the gas is hotter relative to
the group luminosity. One possibility is that the incidence of
mergers, and/or the type of mergers that the two subsamples of groups
have undergone, is different for the two subsamples.

Recent simulations by \citet{row04} find that in {\it major} mergers
(defined as increasing the cluster mass by $\geq 20$ percent) clusters
become brighter and heat up roughly parallel to the $L_{X}/T_{X}$
relation, whereas in minor mergers, the temperature increases and the
luminosity decreases. Although these simulations are for more massive
clusters, this suggests a possible interpretation of our results. If
the radio-loud groups had recently undergone, or are continuing to
undergo, minor mergers, then their $L_{X}/T_{X}$ properties could be
explained. In this model, the radio-quiet groups must either have
undergone mainly major mergers, or be in more isolated environments
where mergers are less common. One possibility is that radio-quiet
groups could be more evolved, so that most merging has already
occurred, and the gas has cooled back to the $L_{X}/T_{X}$ line. In
this scenario, it might be expected that the radio-quiet groups would
resemble fossil groups where mergers are no longer common.

Such a model of group evolution should be testable using the optical
properties of the group. However, we found no evidence for a
difference in either the number of galaxies in the group or the degree
of dominance by the central galaxy for the radio-quiet and radio-loud
groups (Section~\ref{sec:opt}). A model of this sort cannot be ruled
out, as mergers remain the most plausible model for how radio galaxies
are triggered. However, there does not appear to be any evidence that
the radio-quiet and radio-loud groups are in different stages of
evolution, based on the comparison of the optical group properties. We
therefore conclude that radio-source heating is the most plausible of
the three explanations for our results.

\subsection{Models for radio-source heating}
\label{sec:models}
If the radio-source heating model is adopted, then it is necessary to
consider whether it is possible to explain the observed results for
all of the radio-loud groups in the context of this model. In
Fig.~\ref{lt}, it is apparent that there are some ``radio-loud''
groups that do not show a temperature excess, and one particularly
anomalous group that is much cooler than predicted (NGC 3557). In
addition to the sources with no large temperature excess, most of the
currently active radio sources in the sample are unlikely to be
capable of producing all of the heating that is observed. The two
groups HCG~62 and NGC~4636, which show strong evidence for
radio-related structure and a large heating effect (especially in the
latter), are particularly problematic. Either there is low-frequency
radio structure that has not yet been observed, which indicates active
or recently switched-off large-scale radio jets and lobes, or else the
heating effects are long-lived. Some of the most powerful radio
sources, such as 3C~66B, 3C~449, and NGC~4261, are probably capable of
producing the heating that is observed \citep{c03b}, but the heating
in many of the radio-loud groups must be longer-lived than the radio
source.

It is also interesting to consider the mechanisms for heating in
different stages of radio-source evolution. In \citet{c03b}, we argued
that large, powerful FRIs are subsonic and likely to be heating their
surroundings gently via $P$d$V$ work as the lobes expand.  However, in
the early stages of FRI evolution, the sources are known to be
overpressured (even assuming minimum energy pressures), and the recent
observation of a heated shell around the inner lobe of Cen A shows
that shock-heating is not only likely to be an important mechanism in
FRIIs, but plays a role in the early stages of FRI evolution as
well. This process may also be occurring in one of the groups in this
sample, NGC~1052 (see Section~\ref{sec:1052}).  Finally, the {\it
Chandra} observations of NGC~4636 show that additional mechanisms for
AGN heating may exist, as it is difficult to explain the morphology of
the shocked arms of gas via radio-lobe expansion. We conclude that the
heating effects found in the study of the {\it ROSAT} sample presented
here are likely to be the result of different types of radio-source
heating, so that one simple model for the entire sample is unlikely to
be correct. In some sources there may be small regions of shocked gas,
unresolved in the {\it ROSAT} data, leading to the temperature
increases, whereas in others more widespread heating is necessary. A
detailed analysis of {\it XMM} observations, which now exist for a
significant number of the radio-loud groups, would help to investigate
these possibilities, as would a low-frequency radio study to constrain
the properties of the radio sources.

\subsection{The importance of radio galaxies in structure formation
  models}
\label{sec:sfimpl}

We have shown that radio galaxies at some stage of development are
present in up to 50 percent of elliptical-dominated groups
(Section~\ref{sec:parentpop}), and have also presented evidence that
radio-source heating is common. It is therefore of interest to
consider whether their energy input is of significance in the context
of structure formation models. We carried out some simple calculations
to determine whether the energy input from low-power radio sources in
elliptical-dominated groups could be important in the context of the
energy-injection requirements of structure formation models.

We estimated the energy input rate from the average radio source in
the sample by taking the average value of $L_{1.4}$ for the 19
radio-loud groups ($3.5 \times 10^{23}$ W Hz$^{-1}$) and scaling the
kinetic luminosity of 3C~31 (from the model of \citealt{lb02a}) by the
ratio of $L_{1.4}$ for the average radio source and 3C~31, which gives
$7 \times 10^{35}$ W. We then assumed that 1/3 of the kinetic
luminosity is transferred to the group gas, a conservative lower
limit. We assumed a duty cycle of 50 percent, based on the fraction of
radio sources in the Garcia catalogue
(Section~\ref{sec:parentpop}). This gives a typical energy input rate
of $2.3 \times 10^{51}$ keV/s over the lifetime of the group. The
energy rate per particle was determined using the average number of
particles in the group (determined as part of the heat capacity
calculations in Section~\ref{sec:rcorrel}) to be $6.6 \times 10^{-18}$
keV/particle/s. Assuming the injection energy is required to be of
order 1 keV/particle \citep[e.g.][]{wu00}, then the average radio
source in the sample can provide the necessary energy input over $\sim
5 \times 10^{9}$ years, which is a plausible group lifetime (this is
about 10 times the standard radio-galaxy lifetime, so that the radio
source must have $\sim$ 5 active phases during the group lifetime). We
conclude that low-power radio sources may be capable of providing the
necessary energy input in elliptical-dominated groups.  As it is the
elliptical-dominated groups that principally determine the
$L_{X}/T_{X}$ relation for groups, since they dominate the population
of groups with luminous, group-scale X-ray environments, it is
therefore possible that the energy input from {\it low-power} radio
galaxies can explain the X-ray properties of groups.

Other workers have carried out calculations of the energy input of
radio galaxies into clusters. It seems likely that low-power radio
galaxies can provide sufficient energy on average to balance cooling
flows \citep[e.g.][]{fab03}; however, to explain the cluster
$L_{X}/T_{X}$ relation may require more energy than can be supplied by
FRIs [although \citet{roy04} found that effervescent heating by rising
bubbles could solve the entropy problem by heating at large radii].
The energy contribution from FRII radio galaxies and quasars is also
likely to be important in the rarer situations where they occur.

\section{Conclusions}

We have presented a detailed study of the gas properties of
radio-quiet and radio-loud elliptical-dominated groups based on the
{\it ROSAT} groups sample of Osmond \& Ponman (2004). We reach the
following conclusions:
\begin{itemize}
\item 63 percent of the elliptical-dominated groups (19/30) in the OP04
  sample have associated radio sources at the centre of a dominant
  group galaxy.
\item Our sample of elliptical-dominated groups is not significantly
biased in its radio-loud fraction: the true fraction in the
parent population may be $\sim 40 - 50$ per cent.
\item Radio-loud groups are likely to have a higher temperature than
  radio-quiet groups of the same luminosity.
\item The energy required to produce the observed temperature excess
  correlates weakly with the 1.4-GHz radio luminosity of the sources.
\item The difference in gas properties for radio-quiet and radio-loud
  groups is most plausibly interpreted as evidence for radio-source heating.
\item Evidence for radio-source interactions with the surrounding gas
  is found in {\it Chandra} or {\it XMM-Newton} observations of many
  of the radio-loud groups, although there are also several groups
  that show disturbances not directly related to observable radio
  structure.
\item The radio-loud groups are at different stages in the heating
  process, so that some may be experiencing shock-heating by young
  radio sources, some are being gently heated by a currently active
  large radio galaxy, and some show longer-lived heating effects from
  a previous generation of radio-source activity.
\end{itemize}

\section*{Acknowledgments}

We are grateful to Trevor Ponman for providing access to the GEMS
X-ray analysis in advance of publication, and for helpful comments in
the course of JHC's PhD viva. We would also like to thank Diana
Worrall for comments on an early draft, and to thank the referee for helpful suggestions. JHC thanks PPARC for a
studentship. MJH thanks the Royal Society for a research
fellowship. This work made use of the NASA/IPAC Extragalactic Database
(NED), which is operated by the Jet Propulsion Laboratory, California
Institute of Technology, under contract with the National Aeronautics
and Space Administration.

\bibliography{jhc_refs}

\end{document}